\documentclass[letterpaper, 10 pt, conference]{ieeeconf}  

\IEEEoverridecommandlockouts                              
\usepackage[utf8]{inputenc}
\usepackage[T1]{fontenc}

\title{\LARGE \bf
Revolution-Spaced Output-Feedback Model Predictive Control for Station Keeping on Near-Rectilinear Halo Orbits}

\author{Yuri Shimane$^{1}$, Stefano Di Cairano$^{2}$, Koki Ho$^{3}$, and Avishai Weiss$^{4}$
\thanks{$^{1,3}$Y. Shimane and K. Ho are with the Daniel Guggenheim School of Aerospace Engineering, Georgia Institute of Technology, Atlanta, GA 30332, USA
        Emails: {\tt\small \{yuri.shimane,kokiho\} at gatech.edu}}%
\thanks{$^{2,4}$S. Di Cairano and A. Weiss are with Mitsubishi Electric Research Laboratories (MERL), Cambridge, MA 02139, USA
        Emails: {\tt\small \{dicairano,weiss\} at merl.com}}%
}

\usepackage{graphicx}
\usepackage{siunitx}
\usepackage{algorithm}
\usepackage{algpseudocode}
\usepackage{amsfonts}
\usepackage{booktabs}
\usepackage{amsmath}
\usepackage{amssymb}
\usepackage{balance}                        
\usepackage[flushleft]{threeparttable}
\usepackage[font=footnotesize,labelformat=simple]{subcaption}
\newtheorem{proposition}{Proposition}
\newcommand{\EMrot}{\mathrm{EM}}
\newcommand{\Frame}{\mathcal{F}}

\newcommand{\Tbold}{\boldsymbol{T}}
\newcommand{\Pbold}{\boldsymbol{P}}
\newcommand{\Qbold}{\boldsymbol{Q}}
\newcommand{\Rbold}{\boldsymbol{R}}
\newcommand{\Sbold}{\boldsymbol{S}}
\newcommand{\Hbold}{\boldsymbol{H}}
\newcommand{\Lbold}{\boldsymbol{L}}
\newcommand{\Vbold}{\boldsymbol{V}}
\newcommand{\Xbold}{\boldsymbol{X}}
\newcommand{\abold}{\boldsymbol{a}}
\newcommand{\cbold}{\boldsymbol{c}}
\newcommand{\dbold}{\boldsymbol{d}}
\newcommand{\rbold}{\boldsymbol{r}}
\newcommand{\vbold}{\boldsymbol{v}}
\newcommand{\ubold}{\boldsymbol{u}}
\newcommand{\Ubold}{\boldsymbol{U}}
\newcommand{\hbold}{\boldsymbol{h}}
\newcommand{\xbold}{\boldsymbol{x}}
\newcommand{\ybold}{\boldsymbol{y}}
\newcommand{\fbold}{\boldsymbol{f}}
\newcommand{\Ibold}{\boldsymbol{I}}
\newcommand{\Phibold}{\boldsymbol{\Phi}}

\newcommand{\period}{T}

\newcommand{\Nu}{N}

\newcommand{\red}[1]{\textcolor{black}{#1}}
\newcommand{\ACCrev}[1]{\textcolor{black}{#1}}
\newcommand{\review}[1]{\textcolor{black}{#1}}

\begin{document}

\maketitle
\thispagestyle{empty}
\pagestyle{empty}

\begin{abstract}
We develop a model predictive control (MPC) policy for station keeping on a Near-Rectilinear Halo Orbit (NRHO). 
The proposed policy achieves full-state tracking of a reference NRHO via a multiple-maneuver control horizon, each spaced one revolution apart to abide by typical mission operation requirements.
\review{We prove that the proposed policy is recursively feasible, and perform numerical} evaluation in an output-feedback setting by incorporating a navigation filter and realistic operational uncertainties, where the proposed MPC is compared against the state-of-the-art station-keeping algorithm adopted for the Gateway.
\review{Our approach successfully maintains the spacecraft in the vicinity of the reference NRHO at a similar cumulative cost as existing station keeping methods without encountering phase deviation issues, a common drawback of existing methods with one maneuver per revolution.}
\end{abstract}

\section{Introduction}
\begin{figure}
    \centering
    \includegraphics[width=0.85\linewidth]{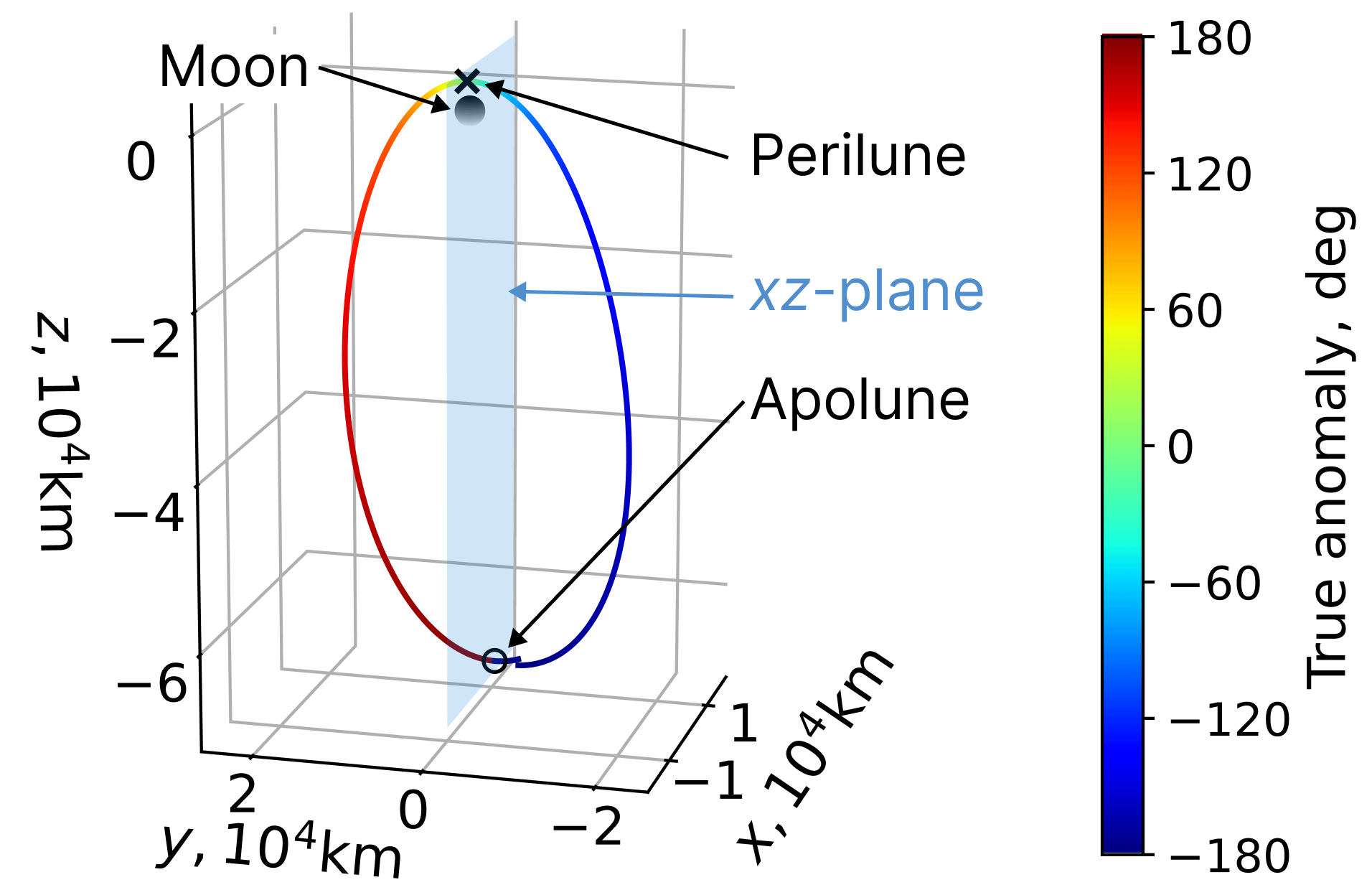}
    \caption{NRHO in Earth-Moon rotating frame}
    \label{fig:NRHO_anatomy}
\end{figure}

With growing interest in lunar exploration, \review{\textit{libration point orbits} (LPOs), quasi-periodic orbits about the equilibrium points of the Earth-Moon-spacecraft three-body system,} offer unique locations to place both robotic and crewed spacecraft.
For example, the lunar Gateway is planned in the 9:2 resonant southern Near-Rectilinear Halo Orbit (NRHO) about the Earth-Moon L2 point~\cite{Lee2019,ZimovanSpreen2022}, shown for one revolution in Figure~\ref{fig:NRHO_anatomy}.
The instability of LPOs requires that the spacecraft conduct station-keeping (SK) maneuvers.
The purpose of SK is to maintain the spacecraft near a pre-computed reference LPO, or \textit{baseline}, in the presence of uncertainties such as state estimation error, modeling error, and control execution error.
Due to the stringent propellant budget, typically higher instability of LPOs compared to traditional orbits around planets and moons, and the low number of heritage missions flying on LPOs, SK techniques on LPOs are an active area of research.

To accommodate mission operations, SK maneuvers are typically required to be as infrequent as possible~\cite{Davis2022}.
On the NRHO with an orbital period of about $6.55$~days, a typical requirement is for SK maneuvers to be conducted at most once every revolution about the Moon.
To adhere to this requirement, a commonly adopted approach is known as \textit{$x$-axis crossing control} (XAC)~\cite{Davis2022}, a shooting-based method for designing SK maneuvers. 
Recently, the CAPSTONE mission~\cite{Cheetham2022} adopted XAC, and variants of XAC are currently being studied for the upcoming Gateway mission~\cite{Davis2022}.

One drawback of XAC stems from the fact that at most three out of the six translational state components can be assigned.
To overcome this deficiency, \review{XAC leverages the LPO's symmetry about the $xz$-plane in the Earth-Moon rotating frame~\cite{Howell1984}, see Figure~\ref{fig:NRHO_anatomy}}.
A subset of the predicted spacecraft state at the intersection with the $xz$-plane is matched with the corresponding state components along the baseline when it intersects the same plane.
Targeting based on $xz$-plane intersections results in a discrepancy between the epoch when the spacecraft crosses the plane and when the baseline crosses the plane.
\review{As a consequence, the steered path may experience a \textit{phase angle deviation}: the spacecraft's location along the orbit may drift ahead or behind the baseline along the same orbit.}
To date, the phase deviation has been treated by ad-hoc heuristics, e.g., augmenting the targeting scheme with the epoch at which the symmetry event occurs~\cite{Davis2022,Williams2023}, or encapsulating the targeting scheme within a constrained optimization problem~\cite{Shimane2024PCSCOP}.

\review{Beyond XAC, a number of model predictive control (MPC) policies have been proposed for SK~\cite{Misra2018,Elango2022Eigenmotion,Quartullo2023,Padhi2024}.
These policies do not suffer from phase angle deviation issues as they provide full-state tracking by considering either multiple maneuvers per revolution~\cite{Elango2022Eigenmotion,Padhi2024} or a continuously controllable spacecraft~\cite{Misra2018,Quartullo2023}.
Moreover, the frequent control actions render these policies unfit for implementation in an actual mission.
Some MPC approaches adopt quadratic cost functions~\cite{Misra2018,Padhi2024} that do not optimize the actual fuel, which is a scarce resource for a spacecraft.}
For further details, see~\cite{Shirobokov2017} and references therein.

In this work, we propose \review{an operationally compliant} MPC policy that overcomes the phase disparity via full-state targeting.
The proposed MPC uses a control horizon with \review{at least} two maneuvers spaced one revolution apart, which provides sufficient controllability to track all six state components.
Simultaneously, the one-revolution control cadence ensures that our approach is consistent with the operational requirement of conducting, at worst, a single SK maneuver per revolution.
To explicitly minimize the propellant consumption, we employ an economic objective~\cite{Rawlings2012,Angeli2015} based solely on the control cost, \review{resulting in similar SK cost to XAC over extended durations}.
The proposed MPC, hereafter denoted as SKMPC, sequentially solves a second-order cone program (SOCP) that steers the state of the spacecraft to the vicinity of the baseline at the end of its targeting horizon.
At each iteration, the SOCP is re-instantiated by linearizing the dynamics about the steered state from the previous iteration; the SKMPC is terminated when the steered state propagated with the nonlinear dynamics lies sufficiently close to the baseline. 
We provide a brief discussion on the recursive feasibility of the SKMPC and numerically demonstrate its performance, \review{with comparison to XAC}.

This work extends~\cite{ACC2025trackingMPC} by incorporating a navigation filter to estimate the full state of the spacecraft, validating the proposed approach in a realistic output-feedback scenario. 
Our simulation incorporates disturbances due to navigational uncertainty, dynamics modeling errors, control actuation errors, and random impulses imparted at scheduled times along the NRHO due to momentum wheel desaturation maneuvers.
We also implement XAC and compare its performance against the SKMPC.
We provide comprehensive Monte Carlo results with varying disturbance levels, thereby quantifying the coupled performance of the filter and the controller.

\section{Background}
\label{sec:background}

\subsection{Spacecraft Dynamics Model}
\red{We consider the spacecraft's motion} in the J2000 inertial frame $\mathcal{F}_{\rm Inr}$~\cite{Acton2018}, centered at the Moon.
The state of the spacecraft $\xbold \in \mathbb{R}^6$ consists of the Cartesian position $\rbold \in \mathbb{R}^3$ with respect to the Moon \red{and} the rate of change of $\rbold$ in $\mathcal{F}_{\rm Inr}$, denoted by $\vbold \in \mathbb{R}^3$. 
The equations of motion are~\cite{Vallado2001}
\begin{equation}
    \dot{\xbold} = \fbold\left(\xbold,t\right)
    = 
    \begin{bmatrix}
        \vbold \\ 
        -\dfrac{\mu}{r^3}\rbold + \abold_{\rm J2} + \sum_{i} \abold_{N_i} + \abold_{\rm SRP}
    \end{bmatrix},
    \label{eq:nonlinear_dynamics}
\end{equation}
where $r = \|\rbold\|_2$, and $\mu$ is the gravitational parameter of the Moon.  
The derivative of $\vbold$ consists, in order, of the Keplerian acceleration due to the Moon, J2 perturbation of the Moon $\abold_{\rm J2}$, gravitational perturbations by other celestial bodies $\abold_{N_i}$, and the solar radiation pressure (SRP) $\abold_{\rm SRP}$, all in $\Frame_{\rm Inr}$.
We include third-body perturbations due to the Earth and the Sun.
\review{For expressions of $\abold_{\rm J2}$, $\abold_{N_i}$, and $\abold_{\rm SRP}$, see~\cite{Vallado2001}.} 
Note that  $\abold_{\rm J2}$, $\abold_{N_i}$ and $\abold_{\rm SRP}$ in equation~\eqref{eq:nonlinear_dynamics} are time-dependent, making $\fbold$ non-autonomous. 
Constants in the equations of motion and ephemerides of celestial bodies are obtained from the SPICE toolkit~\cite{Acton2018}. 

An initial perturbation $\delta \xbold_k$ \review{at time $t_k$} can be linearly propagated to time $t_{k+1}$, denoted by $\delta \xbold_{k+1}$, via the linear state-transition matrix (STM) $\Phibold(t_{k+1},t_k) \in \mathbb{R}^6$, 
\begin{equation}
    \delta \xbold_{k+1} = \Phibold(t_{k+1},t_k) \delta \xbold_k
    .
    \label{eq:delta_x_map_with_STM}
\end{equation}
The Jacobian of the dynamics may be used to construct the STM by solving the matrix initial value problem (IVP)
\begin{equation}    \label{eq:stm_matrix_ivp}
\begin{aligned}
    \dot{\Phibold}(t,t_k) &= \dfrac{\partial \fbold(\xbold,t)}{\partial \xbold}\Phibold(t,t_k)
    ,\quad
    \Phibold(t_k,t_k) = \Ibold_6. 
\end{aligned}
\end{equation}
We use the shorthand notations $\xbold_k = \xbold(t_k)$ and $\Phibold_{j,i} = \Phibold(t_j,t_i)$, \red{and} we express the block submatrices of $\Phibold_{j,i}$ as
\begin{equation}    \nonumber
    \Phibold_{j,i} =
    \begin{bmatrix}
        \Phibold_{j,i}^{\rbold \rbold} & 
        \Phibold_{j,i}^{\rbold \vbold} \\[0.1em]
        \Phibold_{j,i}^{\vbold \rbold} & 
        \Phibold_{j,i}^{\vbold \vbold} \\
    \end{bmatrix}.
\end{equation}

Assuming the control on the spacecraft results in an impulsive change in velocity\footnote{Due to control executions lasting on the order of seconds to minutes along an orbit with a period on the order of days, all conventional thrusters are effectively impulsive in this application~\cite{Vallado2001}.},
\ACCrev{the state at time $t_{k+1}$ following an impulse applied at time $t_k$ is given by
\begin{equation}    \label{eq:nonlinearControlDynamics}
    \begin{aligned}
        \xbold_{k+1} &= 
        \xbold_k + \int_{t_k}^{t_{k+1}} \fbold\left( \xbold, t\right)
        + \delta(t - t_k) \begin{bmatrix}
            \boldsymbol{0}_{3 \times 1} \\ \ubold_{k}
        \end{bmatrix}
        \mathrm{d}t
        ,
    \end{aligned}
\end{equation}
where the control $\ubold_k \in \mathbb{R}^3$ is the impulsive change in velocity and $\delta$ is the Dirac delta function.}

\subsection{Earth-Moon Rotating Frame}
\review{The Earth-Moon rotating frame, $\mathcal{F}_{\rm EM}$, is practical for analyzing the natural motion and devise SK controllers due to the NRHO's symmetry about the frame's $xz$-plane, as illustrated in Figure~\ref{fig:NRHO_anatomy}.
In $\mathcal{F}_{\rm EM}$, the $x$-axis is aligned with the instantaneous Earth-Moon vector, the $z$-axis is aligned with the co-rotating angular velocity vector of the Earth and the Moon, and the $y$-axis completes the triad.}
\review{The matrix $\Sbold^{\rm Inr}_{\rm EM} \in\mathbb{R}^{6\times6}$ that transforms a state from $\Frame_{\rm Inr}$ to $\Frame_{\rm EM}$ is
\begin{equation}
    \Sbold^{\rm Inr}_{\rm EM}(t)
    =
    \begin{bmatrix}
        \Tbold^{\rm Inr}_{\rm EM}(t) & \boldsymbol{0}_{3\times 3} \\
        \dot{\Tbold}^{\rm Inr}_{\rm EM}(t) & \Tbold^{\rm Inr}_{\rm EM}(t)
    \end{bmatrix}
    ,
    \Tbold^{\rm Inr}_{\rm EM} = \begin{bmatrix}
        \boldsymbol{e}_1^T(t) \\[0.1em]
        \boldsymbol{e}_2^T(t) \\[0.1em]
        \boldsymbol{e}_3^T(t)
    \end{bmatrix}
    ,
\end{equation}
where $\boldsymbol{e}_1(t)$, $\boldsymbol{e}_2(t)$, and $\boldsymbol{e}_3(t)$ are the basis vectors of $\Frame_{\rm EM}$ realized in $\Frame_{\rm Inr}$,
\begin{equation}    \nonumber 
    \boldsymbol{e}_1(t) = \dfrac{-\dbold_{\rm Earth}(t)}{\| \dbold_{\rm Earth}(t) \|_2},
    \boldsymbol{e}_3(t) = \dfrac{\dbold_{\rm Earth}(t) \times \vbold_{\rm Earth}(t)}{\| \dbold_{\rm Earth}(t) \times \vbold_{\rm Earth}(t) \|_2}
    ,
\end{equation}
and $\boldsymbol{e}_2(t) = \boldsymbol{e}_3(t) \times \boldsymbol{e}_1(t)$, $\vbold_{\rm Earth}(t)$ is the Earth's velocity with respect to the Moon, and $\dot{\Tbold}^{\rm Inr}_{\rm EM}(t)$ is the time-derivative of $\Tbold^{\rm Inr}_{\rm EM}(t)$ in~$\Frame_{\rm Inr}$.}

\subsection{Stability on Near Rectilinear Halo Orbit}
The NRHO's unstable subspace~\cite{ZimovanSpreen2022} necessitates SK actions to prevent the spacecraft from diverging from the baseline in the presence of uncertainties.
Let $\theta$ denote the \textit{osculating true anomaly} with respect to the Moon~\cite{Vallado2001},
\begin{equation}    \nonumber
    \theta = \operatorname{atan2}\left(
        h v_r, h^2/r - \mu
    \right)
    ,\,
    h = \| \rbold \times \vbold \|_2
    ,\,
    v_r = \frac{\rbold\cdot\vbold}{r}
    .
\end{equation}
\review{As noted in the literature on NRHO~\cite{ZimovanSpreen2022}, the dynamics are most sensitive around \textit{perilune} where $\theta = 0^{\circ}$ and the spacecraft is closest to the Moon, and least sensitive around \textit{apolune} where $\theta = 180^{\circ}$ and the spacecraft is farthest from the Moon.
To make the SK activity as robust as possible against navigation and control actuation errors, SK maneuvers typically execute near apolune.
Let the \textit{maneuver true anomaly} $\theta_{\rm man}$ denote the true anomaly at which the SK maneuver is scheduled to occur.
In accordance with operational plans for the Gateway~\cite{Davis2022}, we use $\theta_{\rm man} = 200^{\circ}$.}

\subsection{Navigation Filter}
We consider an extended Kalman filter (EKF) to estimate the spacecraft state. 
\review{Let $\hat{\xbold} \in \mathbb{R}^6$ and $\Pbold \in \mathbb{R}^{6 \times 6}$ denote the estimate of the state $\xbold$ and the state covariance, respectively.}
We briefly present the prediction and update steps of the EKF, along with the measurement model and the impulse events, \review{where predicted quantities at $t_k$ based on $t_{k-1}$ are denoted by $(\cdot)_{k}^-$, and updated quantities following a measurement or a maneuver at $t_k$ are denoted by $(\cdot)_{k}^+$.} 

\subsubsection{Prediction}
\red{
The prediction step from $t_{k-1}$ to $t_k$ is~\cite{Sarkka2013}
\begin{subequations}
\begin{align}
    \hat{\xbold}_{k}^- &= 
    \hat{\xbold}_{k-1}^+ +
    \int_{t_{k-1}}^{t_k} \fbold \left(\hat{\xbold}, t\right) \mathrm{d}t ,
    \nonumber \\
    \Pbold_{k}^- &= \Phibold_{k,k-1} \Pbold_{k-1}^+ \Phibold_{k,k-1}^T + \Qbold_{k,k-1},
    \nonumber 
\end{align}
\end{subequations}
where \review{$\Phibold_{k,k-1}$ is the STM evaluated along $\hat{\xbold}_{k}^-$}, and $\Qbold_{k,k-1}$ is the process noise accounting for unmodelled disturbances. 
We adopt the unbiased random process noise model~\cite{Carpenter2018}
\begin{equation}
    \Qbold_{k,k-1} = \sigma_p^2 
    \begin{bmatrix}
        ({\Delta t^3}/{3}) \Ibold_{3}
        &
        ({\Delta t^2}/{2}) \Ibold_{3}
        \\
        ({\Delta t^2}/{2}) \Ibold_{3}
        &
        \Delta t \Ibold_{3}
    \end{bmatrix}
    ,
    \,\,
    \Delta t = t_k - t_{k-1}
    \nonumber
\end{equation}
where $\sigma_p$ is a tuning parameter.
}

\subsubsection{Update}
At time $t_k$, a measurement is provided to the filter. 
The noisy measurement $\ybold_k \in \mathbb{R}^m$ is assumed to follow a multivariate normal distribution with zero mean and covariance $\Rbold_k$~\cite{Sarkka2013} \review{such that $\ybold_k = \hbold\left(\hat{\xbold}_k^-\right) + \mathcal{N}(\boldsymbol{0}_{m \times 1}, \Rbold_k)$.
Let $\Hbold_k = \Hbold(\xbold_k) =  \partial \hbold\left(\hat{\xbold}_k^-\right) / \partial \xbold$; the update step is~\cite{Sarkka2013}
\begin{subequations}
\begin{align}
    \Lbold_k &= \Pbold_{k}^- \Hbold_k^T 
    \left( \Hbold_k \Pbold_{k}^- \Hbold_k^T + \Rbold_k \right)^{-1},
    \label{eq:ekf_kalman_gain}
    \nonumber \\
    \hat{\xbold}_{k}^+ &= \hat{\xbold}_{k}^- + \Lbold_k 
    \left( \ybold_k - \hbold_k\left(\hat{\xbold}_{k}^-\right) \right),
    \nonumber \\
    \Pbold_{k}^+ &= (\Ibold_6 - \Lbold_k \Hbold_k) \Pbold_{k}^-(\Ibold_6 - \Lbold_k \Hbold_k)^T 
    + \Lbold_k \Rbold_k \Lbold_k^T.
    \nonumber
\end{align}
\end{subequations}
The Joseph form of the covariance update has been adopted for superior numerical stability~\cite{Carpenter2018}.
}

\subsubsection{Measurements}
We consider measurements based on range and range-rate, with $\hbold$ and $\Hbold$ given by
\begin{subequations}
\begin{align}
    \hbold(\xbold) 
    &= \begin{bmatrix}
        r
        \\
        \dot{r}
    \end{bmatrix} = \begin{bmatrix}
        \| \rbold \|_2 
        \\
        \rbold^T \vbold / \| \rbold \|_2
    \end{bmatrix}
    ,
    \nonumber
    \\[0.3em]
    \Hbold(\xbold) &= 
    \begin{bmatrix}
        \dfrac{x}{r} & \dfrac{y}{r} & \dfrac{z}{r} & 0 & 0 & 0\\
        \dfrac{v_x}{r} - \dfrac{x \dot{r}}{r^2} &
        \dfrac{v_y}{r} - \dfrac{y \dot{r}}{r^2} &
        \dfrac{v_z}{r} - \dfrac{z \dot{r}}{r^2} &
        \dfrac{x}{r} & \dfrac{y}{r} & \dfrac{z}{r}
    \end{bmatrix}
    ,
    \nonumber
\end{align}
\end{subequations}
\review{where $\rbold$, $\vbold$, and the state components $[x,y,z,v_x,v_y,v_z]$ are in $\Frame_{\rm Inr}$.}
We assume a constant measurement covariance $\Rbold_k = \Rbold = \operatorname{diag}(\sigma_r^2, \sigma_{\dot{r}}^2)$
where $\sigma_r$ and $\sigma_{\dot{r}}$ are the standard deviations of the range and range-rate measurements.

\subsubsection{Impulse Events}
We model SK maneuvers as a velocity impulse on the spacecraft $\ubold_k = \hat{\ubold}_k + \mathcal{N} (\boldsymbol{0}_{3\times 1}, \Vbold_k)$, where $\hat{\ubold}$ is the expected impulse, and $\Vbold_k$ is the corresponding covariance. 
The maneuver estimate $\hat{\ubold}$ is computed by the SK controller and $\Vbold_k = (\sigma_{\ubold, \mathrm{abs}} + \sigma_{\ubold, \mathrm{rel}} \|\hat{\ubold}\|_2)^2 \Ibold_3$,
where $\sigma_{\ubold, \mathrm{abs}}$ and $\sigma_{\ubold, \mathrm{rel}}$ are the absolute and relative standard deviation of the thruster.
Then, the state is updated via $\hat{\xbold}_{k}^+ = \hat{\xbold}_{k}^- + [\boldsymbol{0}_{3\times 1}^T, \hat{\ubold}^T]^T$, and the covariance is updated via
\begin{subequations}
\begin{align}
    \Pbold_{k}^+ &= \Pbold_{k}^- +
    \begin{bmatrix}
        \boldsymbol{0}_{3\times 3} & \boldsymbol{0}_{3\times 3} \\
        \boldsymbol{0}_{3\times 3} &
        \Vbold_k
    \end{bmatrix}.
    \nonumber
\end{align}
\end{subequations}
\review{Such an update following an uncertain maneuver effectively works as a partial reset to the EKF.
Had we not assumed an impulsive control model, we would need to consider a compensation model with intensities scaled according to the maneuver magnitude~\cite{Carpenter2018}.}

\subsection{$x$-Axis Crossing Control}
\review{We now provide a brief review of XAC. 
XAC attempts to maintain the symmetry of the controlled trajectory about the $xz$-plane in $\Frame_{\rm EM}$~\cite{Davis2022}.
At a given control time $t_k$, let $t_{Np}$ and $t_{Np,\rm ref}$ denote the $N_{\rm rev}^{\rm th}$ times the predicted trajectory and the baseline cross the $xz$-plane near perilune, respectively, and $\xbold_{Np,\rm ref}$ denote the baseline state at $t_{Np,\rm ref}$.
XAC is a single shooting problem to obtain a control $\ubold_k$ that results in residual $F$ defined as the difference between the actual and the baseline $x$-axis component of the velocity in $\Frame_{\rm EM}$ when crossing the $xz$-plane,
\begin{equation}    \nonumber
    \begin{aligned}
        F &= \boldsymbol{\Lambda}
            \left(\Sbold^{\rm Inr}_{\rm EM}(t_{Np}) \xbold_{Np} - \Sbold^{\rm Inr}_{\rm EM}(t_{Np,\rm ref}) \xbold_{Np,\rm ref} \right)
        ,
        \\
        \boldsymbol{\Lambda} &= \begin{bmatrix} 0 & 0 & 0 & 1 & 0 & 0 \end{bmatrix}
        ,
        \\
        \xbold_{Np} &= \hat{\xbold}_{k}^- + \begin{bmatrix} \boldsymbol{0}_{3\times1}\\ \ubold_k \end{bmatrix} + \int_{t_k}^{t_{Np}} \fbold[\xbold(t),t] \mathrm{d}t
        ,
    \end{aligned}
\end{equation}
being below a user-defined threshold $\epsilon_F$.
Note that $t_{Np} \neq t_{Np,\rm ref}$ since these are the times at the $xz$-plane crossing along two different trajectories.
The maneuver $\ubold(t_k)$ is obtained by Newton-Raphson iterations,
\begin{equation}    \label{eq:minimum_norm_update}
    \begin{aligned}
        \ubold_k &\gets 
        \ubold_k -
        \mathrm{D}F^T
        \left(\mathrm{D}F \, \mathrm{D}F^T \right)^{-1} F\left(\ubold_k\right)
        ,
        \,\,
        \mathrm{D}F = \dfrac{\partial F}{\partial \ubold_k}
        ,
    \end{aligned}
\end{equation}
until $|F| \leq \epsilon_{v_x}$.
Since XAC suffers from phase dispersion, Davis et al.~\cite{Davis2022} propose augmenting $F$ with the difference between $t_{Np}$ and $t_{Np,\rm ref}$; however, this approach requires balancing physically different quantities of $F$ during the Newton-update, making it sensitive to hyperparameters and non-intuitive to tune~\cite{Shimane2024PCSCOP}.}

\section{Full-State Targeting Economic MPC for Station-Keeping on NRHO}
\label{sec:EMPCforSK}
SKMPC computes a sequence of maneuvers based on the state estimate~$\hat{\xbold}_{k}^-$ at the current time $t_k$ to ensure the predicted state at some future time $t_f > t_k$ lies in the vicinity of the baseline.
\review{To reduce the sensitivity of the errors on the targeted state, we select $t_f$ to correspond to the apolune along the baseline approximately $N_{\rm rev}$ revolutions in the future.
This proposed method computes an SK maneuver at the prescribed $\theta_{\rm man}$ such that the steered state lies in the vicinity of the baseline at the $N_{\rm rev}^{\mathrm{th}}$ apolune into the future.}

\subsection{Problem Formulation}
\begin{figure}
    \centering
    \includegraphics[width=0.95\linewidth]{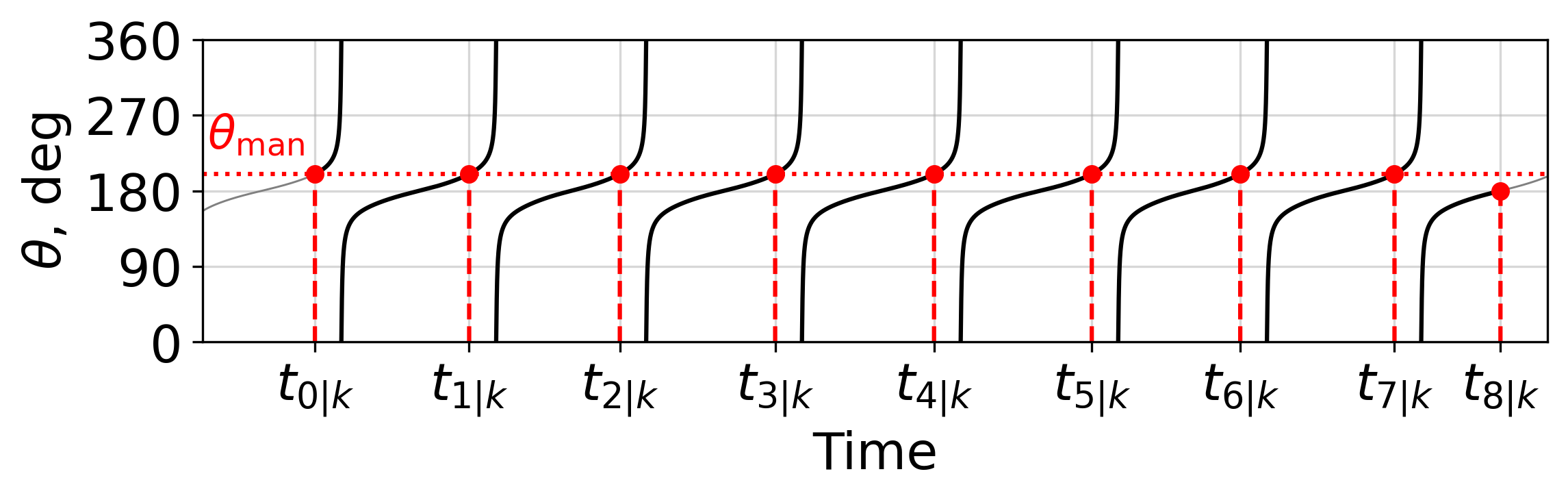}
    \caption{Control horizon with $N=9$ and $\theta_{\rm man} = 200^{\circ}$}
    \label{fig:control_horizon}
\end{figure}

\review{Let $(\cdot)_{j|k}$ denote a quantity predicted at $j$ time increments ahead of the EKF's estimate $\hat{\xbold}_{k}^-$ at $t_k$.
Let $\Nu = N_{\rm rev}+1$ denote the number of impulse maneuvers in the control horizon, with controls denoted by $\ubold_{j|k} \in \mathbb{R}^3$, $j = 0,\ldots,\Nu-1$, occurring at times
\begin{equation}
    \label{eq:control_horizon_definition}
    t_{j|k} =
    \begin{cases}
        t(\theta_{\rm man}, j) & j \leq \Nu-2,
        \\
        t(180^{\circ}, j) & j = \Nu-1,
    \end{cases}
\end{equation}
where $t(a,b)$ denote the earliest time $t \geq t_k$ such that $\theta(t) = a$ after completing $b$ revolutions.
Figure~\ref{fig:control_horizon} illustrates the control horizon defined by~\eqref{eq:control_horizon_definition} for $N=9$ and $\theta_{\rm man} = 200^{\circ}$, corresponding to about $50$~\SI{}{days}.}

We formulate a minimization problem with an economic sum-of-2-norm objective of $\Nu$ maneuvers, which corresponds directly to the propellant mass consumed via Tsiolkovsky's rocket equation \cite{Vallado2001}.
\review{
Let $\Xbold \in \mathbb{R}^{6 \times \Nu}$ and $\Ubold \in \mathbb{R}^{3 \times \Nu}$ denote the concatenated states and controls
\begin{equation} \nonumber 
    \Xbold = \begin{bmatrix}
        \xbold_{0|k} & \cdots & \xbold_{\Nu-1|k}
    \end{bmatrix}
    , \,\,
    \Ubold = \begin{bmatrix}
        \ubold_{0|k} & \cdots & \ubold_{\Nu-1|k}
    \end{bmatrix}
    .
\end{equation}
Let $\bar{\xbold}_{j|k}$ denote the predicted trajectory prior to computing the SK maneuvers, and $\bar{\ubold}_{j|k}$ denote pre-planned maneuvers, initialized to $\boldsymbol{0}_{3 \times 1}$.
We define $\bar{\Xbold}$ and $\bar{\Ubold}$ as}
\begin{equation}    \nonumber
    \bar{\Xbold} = \begin{bmatrix}
        \bar{\xbold}_{0|k} & \cdots & \bar{\xbold}_{\Nu-1|k}
    \end{bmatrix}
    , \,\,
    \bar{\Ubold} = \begin{bmatrix}
        \bar{\ubold}_{0|k} & \cdots & \bar{\ubold}_{\Nu-1|k}
    \end{bmatrix}
    .
\end{equation}

\review{We begin by defining the discrete-time optimal control problem (OCP) assuming linearized dynamics around $\bar{\xbold}$, then will subsequently consider a sequential linearization scheme in Section~\ref{sec:SequentialLinearization}, where the OCP will be solved repeatedly, each time updating $\bar{\Xbold}$ and $\bar{\Ubold}$.
The OCP with linearized dynamics is}
\begin{subequations}
\label{eq:optim_socp_general}
\begin{align}
    \min_{\Xbold,\Ubold}
        \quad& \sum_{j=0}^{\Nu-1} \| \ubold_{j|k} \|_2
    \label{eq:objective}
    \\
    \text{s.t.} \quad &
        \begin{aligned}
            &\xbold_{j+1|k} = \Phibold_{j+1,j|k}\xbold_{j|k} + 
            \begin{bmatrix}
                \Phibold_{j+1,j|k}^{\rbold \vbold} \\[0.1em]
                \Phibold_{j+1,j|k}^{\vbold \vbold}
            \end{bmatrix}
            \ubold_{j|k} + \cbold_{j+1,j|k}
            \\
            &\quad \quad j = 0,\ldots,\Nu-2
        \end{aligned}
        \label{eq:constraint_dynamics}
    \\&
        \xbold_{0|k} = \hat{\xbold}_{k}^-
        \label{eq:constraint_initial_conditions}
    \\&
        \xbold_{\Nu-1|k} + \begin{bmatrix}
            \boldsymbol{0}_{3\times1} \\ \ubold_{N-1|k}
        \end{bmatrix}
        \in \mathcal{X}_{\rm targ}(t_{N-1|k})
    \label{eq:constraint_x_terminal_set}
    \\
    &   \xbold_{j|k} \in \mathcal{X}, \quad \forall j = 0,\ldots,\Nu-1
    \\
    &   \ubold_{j|k} \in \mathcal{U}, \quad \forall j = 0,\ldots,\Nu-1,
    \label{eq:constraint_u_admissible}
\end{align}
\end{subequations}
where $\mathcal{X}$ is the admissible set for the state, $\mathcal{U}$ is the admissible set for the control, \review{$\Phibold_{j+1,j|k}$} is the STM from \review{$t_{j|k}$ to $t_{j+1|k}$} with initial condition \review{$\bar{\xbold}_{j|k}$ and impulsive control $\bar{\ubold}_{j|k}$, and $\cbold_{j+1,j|k}$} is given by
\begin{align}
    \label{eq:linearized_dynamics_c}
    \cbold_{j+1,j|k} &=
    \bar{\xbold}_{j+1,j|k}
    -
    \Phibold_{j+1,j|k}
    \left( \bar{\xbold}_{j|k} +
    \begin{bmatrix}
        \boldsymbol{0}_{3\times1} \\ \bar{\ubold}_{j|k}
    \end{bmatrix}\right)
    ,
    \\  \nonumber
    \bar{\xbold}_{j+1,j|k} &= \bar{\xbold}_{j|k} + \int_{t_{j|k}}^{t_{j+1|k}} \fbold\left(\bar{\xbold}, t\right) 
        + \delta (t-t_k) \begin{bmatrix}
            \boldsymbol{0}_{3\times1} \\ \bar{\ubold}_{j|k}
        \end{bmatrix}
    \mathrm{d}t
    .
\end{align}
The linearized dynamics in~\eqref{eq:constraint_dynamics} implies that the control action $\ubold_{j|k}$ shifts the state within some trust-region $\boldsymbol{\delta} \in \mathbb{R}^6$,
\begin{equation}   \nonumber
    \left|
        \xbold_{j|k} - \bar{\xbold}_{j|k}
    \right| \leq \boldsymbol{\delta}
    ,\quad 
    j - 0,\ldots,N-1
    .
\end{equation}
Constraint~\eqref{eq:constraint_initial_conditions} enforces the initial state to coincide with the current state estimate, and constraint~\eqref{eq:constraint_x_terminal_set} enforces the final state to lie in $\mathcal{X}_{\rm targ}(t_{N-1|k})$.
\red{In~\eqref{eq:constraint_u_admissible}, $\mathcal{U}$} is defined with a maximum executable control magnitude $u_{\max}$ as
\begin{equation}\label{eq:UmaxDefinition}
    \mathcal{U} =
    \left\{
        \ubold \in \mathbb{R}^3 : \|\ubold\|_2 \leq u_{\max}
    \right\}
    .
\end{equation}

\subsection{Definition of Terminal Constraint Set}
Let $\mathcal{X}_{\rm targ}(t_{\Nu-1|k})$ be a 6D ellipsoid centered at the baseline state $\xbold_{f,\mathrm{ref}} \triangleq [\rbold_{f,\mathrm{ref}}^T, \vbold_{f,\mathrm{ref}}^T]^T$ in $\Frame_{\EMrot}$ at time $t_{\Nu-1|k}$,
\begin{equation}
    \label{eq:ellipsoid_definition}
    \begin{aligned}
    &\mathcal{X}_{\rm targ}(t_{\Nu-1|k})
    =
    \\
    &\,\,
    \left\{
        \xbold \in \mathbb{R}^n :
        \| \rbold - {\rbold}_{f,\mathrm{ref}} \|_2 \leq \epsilon_r , 
        \| \vbold - {\vbold}_{f,\mathrm{ref}} \|_2 \leq \epsilon_v
    \right\},
    \end{aligned}
\end{equation}
where $\epsilon_r$ and $\epsilon_v$ are the magnitudes of the apses of the ellipsoid along position and velocity components, and are tuning parameters.
The terminal constraint~\eqref{eq:constraint_x_terminal_set} can be replaced by two second-order cone (SOC) constraints,
\begin{subequations}    \label{eq:termnal_set_ellipsoid}
    \begin{align}
        \left\| \rbold_{\Nu-1|k}^{\rm EM} - \rbold_{f,\mathrm{ref}} \right\|_2 \leq \epsilon_r
        ,
        \\
        \left\| \vbold_{\Nu-1|k}^{\rm EM} + \Tbold^{\rm Inr}_{\rm EM} \ubold_{\Nu-1|k} - \vbold_{f,\mathrm{ref}} \right\|_2 \leq \epsilon_v
        ,
    \end{align}
\end{subequations}
\review{where $\rbold_{\Nu-1|k}^{\rm EM} = \Tbold^{\rm Inr}_{\rm EM}(t_{N-1|k}) \rbold_{\Nu-1|k}$ and $\vbold_{\Nu-1|k}^{\rm EM} = \dot{\Tbold}^{\rm Inr}_{\rm EM}(t_{N-1|k}) \rbold_{\Nu-1|k} + \Tbold^{\rm Inr}_{\rm EM}(t_{N-1|k}) \vbold_{\Nu-1|k}$.}

\subsection{Recursive Feasibility}
\label{sec:theoretical_results}
Next, we briefly discuss the recursive feasibility of problem~\eqref{eq:optim_socp_general} with input constraint~\eqref{eq:UmaxDefinition} and terminal set constraint~\eqref{eq:ellipsoid_definition}. 
The non-autonomous dynamics~\eqref{eq:nonlinear_dynamics} results in the terminal set \review{$\mathcal{X}_{\rm targ}(t_{N-1|k})$} to also be time-dependent, complicating the application of the classical approach for proving recursive feasibility of MPC.
Computing and storing such a time-varying set for the NRHO, which is not periodic but only quasi-periodic, may be prohibitive in practice. 
However, this specific application has some favorable conditions that help us recover guarantees of recursive feasibility.
First, for the considered family of orbits, the STM in~\eqref{eq:constraint_dynamics} ensures controllability of the linearized system around the nominal orbit, described by $\fbold( \xbold_k, t)$. 
\red{Second, the available thrust upper-bounded by~$u_{\max}$ is} ``significantly larger'' than what is required in SK maneuvers, although the general desire is to minimize the requested thrust.

\begin{proposition}
    Let $2 \leq K \leq \Nu$ correspond to the number of maneuvers such that
    \begin{equation*}
        {\rm rank}\left(\Bigg[
        \begin{bmatrix}
            \Phibold_{\Nu-1,0|k}^{\rbold\vbold} \\
            \Phibold_{\Nu-1,0|k}^{\vbold\vbold} 
        \end{bmatrix} \cdots
        \begin{bmatrix}
            \Phibold_{\Nu-1,K-1|k}^{\rbold\vbold} \\
            \Phibold_{\Nu-1,K-1|k}^{\vbold\vbold}
        \end{bmatrix}  \Bigg]  \right) = 6
        .
    \end{equation*}
    For a large enough $u_{\rm max}$, if \eqref{eq:optim_socp_general} is feasible at time $t_{k-1}$ then it is feasible at $t_{k}$. Furthermore, the trajectory remains bounded in a set~$\mathcal{X}_{\rm bnd}$ at the apolune times $t_k$.
\end{proposition}

{\em Proof.}
Since~\eqref{eq:optim_socp_general} is feasible at $t_{k-1}$, there exists 
\review{$U_K(t_{k-1}) = [\ubold_{0|k-1}, \ldots, \ubold_{K-1|k-1}]$} such that
\review{$\xbold_{\Nu-1|k-1}\in\mathcal{X}_{\rm targ}(t_{\Nu-1|k-1})$, where $\xbold_{\Nu-1|k-1}$} is obtained by applying the entire sequence $U_K(t_{k-1})$ via~\eqref{eq:constraint_dynamics} followed by open loop evolution from \review{$t_{K-1|k-1}$ to $t_{\Nu-1|k-1}$}.
Let \review{$\xbold_{\Nu|k-1} = \xbold_{\Nu-1|k-1} + \int_{t_{\Nu-1|k-1}}^{t_{\Nu-1|k}} \fbold\left(\xbold,t\right){\rm d}t$}.
We need to prove that it is possible to obtain a state perturbation $\delta \xbold_{\Nu-1|k}$ such that \review{$\xbold_{\Nu|k-1}+\delta \xbold_{\Nu-1|k}\in \mathcal{X}_{\rm targ}(t_{\Nu-1|k})$.}

Consider the candidate control sequence 
\review{$U_K(t_{k}) = [\ubold_{1|k-1}\!+\!\delta\ubold_{0|k}, \ldots, \ubold_{K-1|k-1}\!+\!\delta\ubold_{K-2|k}, \ubold_{K-1|k}]$} and
\begin{equation*}
    \begin{aligned}
        &\delta \xbold_{\Nu-1|k}
        \\ &
            = \sum_{j=0}^{K-2}
            \begin{bmatrix}
                \Phibold_{\Nu-1,j|k}^{\rbold\vbold} \\
                \Phibold_{\Nu-1,j|k}^{\vbold\vbold} 
            \end{bmatrix}
            \delta \ubold_{j|k}
            +
            \begin{bmatrix}
                \Phibold_{\Nu-1,K-1|k}^{\rbold\vbold} \\
                \Phibold_{\Nu-1,K-1|k}^{\vbold\vbold}
            \end{bmatrix} \ubold_{K-1|k}
            .
    \end{aligned}
\end{equation*}
Then, \review{$\xbold_{\Nu-1|k} = \xbold_{\Nu|k-1} + \delta \xbold_{\Nu-1|k} \in \mathcal{X}_{\rm targ}({t_{\Nu-1|k}})$} is guaranteed by the controllability in $K$ steps for some \review{$\Delta U_K(t_{k}) = [ \delta\ubold_{0|k}, \ldots, \delta\ubold_{K-2|k}, \ubold_{K-1|k} ]$} which perturbs and extends the previous control sequence $U_K(t_{k-1})$.
For a large enough~$u_{\rm max}$, the sequence $ U_K(t_{k}) $ is feasible.
Due to the finite horizon and the bounded thrust, the trajectories remain bounded in a set~$\mathcal{X}_{\rm bnd}$ because the control strategy  enforces~\eqref{eq:constraint_x_terminal_set}, and \review{$\mathcal{X}_{\rm targ}(t_{\Nu-1|k})$} is bounded by~\eqref{eq:ellipsoid_definition}.
\QED

\red{With regards to the assumptions in Proposition 1, due to the quasi-periodic nature of the orbit,} the difference between \review{$\mathcal{X}_{\rm targ}(t_{\Nu-1|k-1})$ and $\mathcal{X}_{\rm targ}(t_{\Nu-1|k})$} is usually small. 
The necessitated correction~$\delta \xbold$ is thus relatively small compared to the control authority~$u_{\max}$.
Hence, the maximum thrust of the propulsion system will be sufficient to ensure the feasibility of the candidate control sequence. 
The controllability of the spacecraft in the NRHO orbit ensures the rank condition.

\review{The economic structure of the SKMPC, along with its operational setting, complicates the theoretical demonstration of closed-loop stability.
Angeli~\cite{Angeli2015} demonstrates closed-loop stability for economic MPC operating with respect to an equilibrium point for a strictly dissipative system by introducing a dissipation inequality.
With the present SKMPC, the control is with respect to a quasi-periodic orbit in non-autonomous dynamics and in the presence of uncertainties.
Guaranteeing closed-loop stability would require modifying the OCP~\eqref{eq:optim_socp_general}, which would compromise its fuel-optimal structure~\cite{rawlings2017model}, and it is not clear whether stability, as opposed to bounded deviation, is desirable in this application due to the impact on fuel consumption.
From an empirical standpoint, the behavior of the SKMPC under uncertainty is demonstrated through numerical simulation in Section~\ref{sec:numerical_results}.}

\subsection{Sequential Linearization Scheme}
\label{sec:SequentialLinearization}
To reduce the prediction error introduced by the linearization in~\eqref{eq:constraint_x_terminal_set}, we employ a sequential linearization scheme~\cite{Shimane2024PCSCOP,Elango2022}, where problem~\eqref{eq:optim_socp_general} is solved sequentially, each time re-linearizing the dynamics.

\review{At each iteration of the sequential linearization, we update $\bar{\Xbold}$ and $\bar{\Ubold}$ with the optimal solution $\Xbold^*$ and $\Ubold^*$ to problem~\eqref{eq:optim_socp_general} and recompute $\Phibold_{j+1,j|k}$ and $\cbold_{j+1,j|k}$ using equations~\eqref{eq:stm_matrix_ivp} and~\eqref{eq:linearized_dynamics_c} along the updated the predicted states $\bar{\xbold}_{j|k}$ and pre-planned maneuvers $\bar{\ubold}_{j|k}$ for $j=0,\ldots,N-1$.}

Algorithm \ref{alg:mpc_tracking} summarizes the SKMPC with the sequential linearization scheme. 
At $t_k$, the algorithm requires as input \review{the control horizon $t_{0|k},\ldots,t_{N-1|k}$, the current state estimate $\hat{\xbold}_k^-$, the baseline state history $\xbold_{\rm ref}(t)$}, terminal constraint set $\mathcal{X}_{\rm targ}(t_{N-1|k})$, admissible control set $\mathcal{U}$, and the maximum number of iterations for linearization $M$.
\review{In line 1, $\bar{\Xbold}$ is initialized using the current state estimate $\hat{\xbold}_k^-$ and the baseline state at $t_{j|k}$ for $j = 1,\ldots,N-1$, denoted by $\xbold_{\rm ref}(t_{j|k})$. Note that solving~\eqref{eq:optim_socp_general} does not require a dynamically feasible initial sequence $\bar{\xbold}_{j|k}$.
In line 4, \texttt{linearize} constructs the linearized expressions $\Phibold_{j+1,j|k}$ and $\cbold_{j+1,j|k}$ around the a priori solution $\bar{\xbold}_{j|k}$ and $\bar{\ubold}_{j|k}$ for $j=0,\ldots,N-1$.
In line 5, \texttt{SOCP} solves~\eqref{eq:optim_socp_general} with a convex solver, e.g., here we use the interior point solver Clarabel~\cite{Clarabel_2024}.}
\review{The algorithm terminates once the solution $\Xbold^*$ and $\Ubold^*$ to~\eqref{eq:optim_socp_general} satisfies the nonlinear dynamics~\eqref{eq:nonlinearControlDynamics} for $j=0,\ldots,N-2$}, and returns the earliest control $\ubold_{0|k}$ for execution.
Then, the spacecraft remains in the corrected orbit until the next maneuver instance $t_{k+1}$, at which time Algorithm~\ref{alg:mpc_tracking} is called again to compute $\ubold_{0|k+1},\ldots,\ubold_{N-1|k+1}$.
\begin{algorithm}
    \caption{SKMPC with Sequential Linearization}
    \label{alg:mpc_tracking}
    \textbf{Inputs}: $t_{0|k}, \ldots, t_{N-1|k}$, $\hat{\xbold}_{k}^-$, $\xbold_{\rm ref}(t)$ $\mathcal{X}_{\rm targ}(t_{N-1|k})$, $\mathcal{U}$, $M$
    \begin{algorithmic}[1]
        \State $\bar{\Xbold} \gets \begin{bmatrix}
            \hat{\xbold}_k^- & \xbold_{\rm ref}(t_{1|k}) & \cdots & \xbold_{\rm ref}(t_{N-1|k})
        \end{bmatrix}$
        \State ${\bar{\Ubold}} \gets \boldsymbol{0}_{3 \times N}$
        \While{true}
            \State $\Phibold, \cbold \gets$\verb|linearize|$(\bar{\Xbold}, \bar{\Ubold})$
            \State $\Xbold^*,\Ubold^* \gets$\verb|SOCP|$(\hat{\xbold}_{k}^-,\Phibold, \cbold,\mathcal{X}_{\rm targ}(t_{N-1|k}),\mathcal{U})$
            \If{\eqref{eq:nonlinearControlDynamics} is satisfied for $j=0,\ldots,\Nu-2$}
                \State break
            \EndIf
            \State $\bar{\Xbold}, \bar{\Ubold} \gets \Xbold^*, \Ubold^*$
            \Comment{Update for next iteration}
        \EndWhile
    \end{algorithmic}
    \textbf{Outputs}: $\ubold_{0|k}^*$
\end{algorithm}

\section{Simulation Setup}
\label{sec:experiment_setup}
The SKMPC is tested on a realistic SK scenario for a spacecraft flying on the NRHO.
The simulation consists of recursively applying Algorithm~\ref{alg:mpc_tracking} for an extended number of revolutions spanning multiple years, subject to navigation errors from the EKF, as well as modeling errors, control execution errors, and impulses due to momentum wheel desaturation. 

\subsection{Error Models}
\red{
The filter is initialized with initial covariance $\Pbold_{0|0} = \operatorname{diag}([\sigma_{r_0}^2,\sigma_{r_0}^2,\sigma_{r_0}^2,\sigma_{v_0}^2,\sigma_{v_0}^2,\sigma_{v_0}^2])$ and initial state estimate 
\begin{equation}
\begin{aligned}
    \hat{\xbold}_{0}^+ &= \xbold(t_{0}) + \mathcal{N}(\boldsymbol{0}_{6 \times 1}, \Pbold_{0|0})
    ,
    \\
    \xbold(t_{0}) &= \xbold_{\rm ref}(t_{0}) + \mathcal{N}\left(\boldsymbol{0}_{6 \times 1}, \operatorname{block-diag}(\sigma_{r_0}^2 \Ibold_3, \sigma_{v_0}^2 \Ibold_3)
    \right)
    ,
\end{aligned}
    \nonumber
\end{equation}
where $\xbold(t_{0})$ is the true state at the initial epoch $t_{0}$, \review{$\xbold_{\rm ref}(t_{0})$ is the baseline state at $t_0$, $\sigma_{r_0}$ is the initial position standard deviation, and $\sigma_{v_0}$ is the initial velocity standard deviation.}
At each revolution, when the spacecraft arrives at $\theta(t) = \theta_{\rm man} = 200^{\circ}$, a maneuver is computed using the predicted state of the filter, $\hat{\xbold}_{k}^-$ and a control horizon defined by~\eqref{eq:control_horizon_definition}.
The true state of the spacecraft is imparted with a corrupted maneuver using the Gates model~\cite{Gates1963}. 
We incorporate dynamics errors, which consist of variation in SRP magnitude and random impulses due to momentum wheel desaturation~\cite{Davis2022}. 
The former is modeled by relative perturbations $\delta (A/m)$ and $\delta C_r$ on $A/m$ and $C_r$ in the $\abold_{\rm SRP}$ term of~\eqref{eq:nonlinear_dynamics}, and the latter is modeled by an additive velocity perturbation $\Delta \vbold$ with random direction and magnitude when the spacecraft arrives at $\theta(t) = \theta_{\rm desat}$, where $\theta_{\rm desat}$ are desaturation true anomalies dictated by mission requirements~\cite{Davis2022}.}

Table~\ref{tab:error_parameters} summarizes the error parameters, corresponding to the assumed levels of uncertainties for the Gateway \cite{Davis2022}, \ACCrev{along with the selected process noise parameter $\sigma_p$}.
\ACCrev{Note that the choice of $\sigma_p$ is dependent on the canonical scales in which the dynamics are expressed.}

\begin{table}[ht]
    \caption{Simulation parameters}
    \label{tab:error_parameters}
    \centering
    \begin{tabular}{ll}
        \toprule
        Simulation parameter & Value  \\
        \midrule
        Average SRP $A/m$, \SI{}{m^2/kg} & $315/17900$ \\
        Average SRP $C_r$ & $2$ \\
        SRP rel. $\delta (A/m)$ 3-$\sigma_{A/m}$, \% & $30$ \\
        SRP rel. $\delta C_r$ 3-$\sigma_{C_r}$, \% & $15$ \\
        Desaturation velocity magnitude 3-$\sigma_{\rm desat}$, \SI{}{cm/s} & $1.0$ \\
        Desaturation true anomaly $\theta_{\rm desat}$, \SI{}{deg} & 
            \begin{tabular}[c]{@{}l@{}}$0^{\circ}$ or\\$330^{\circ}, 0^{\circ}$ or\\$330^{\circ}, 0^{\circ}, 30^{\circ}$\end{tabular}
            \\
        Maneuver rel. magnitude error 3-$\sigma_{\ubold,\rm rel}$, \% & $1.5$ \\
        Maneuver abs. magnitude error 3-$\sigma_{\ubold,\rm abs}$, \SI{}{mm/s} & $1.42$ \\
        Maneuver execution direction error 3-$\sigma_{\Delta \vbold,\rm dir}$, \SI{}{deg} & $1^{\circ}$ \\
        Initial position standard deviation 3-$\sigma_{r_0}$, \SI{}{km} & $10$ \\
        Initial velocity standard deviation 3-$\sigma_{v_0}$, \SI{}{mm/s} & $10$ \\
        Range measurement 3-$\sigma_r$, \SI{}{m} &   $1$ \\
        Range-rate measurement 3-$\sigma_{\dot{r}}$, \SI{}{mm/s} &   $0.1$ \\
        Process noise parameter $\sigma_p$ & $5 \times 10^{-5}$ \\
        \bottomrule
    \end{tabular}
\end{table}

\subsection{Navigation Update Model}
In accordance with the typical operation of ground-based tracking, we assume measurements are provided during \textit{tracking windows}, each lasting $\Delta t_{\rm track} = 1$ hour. 
Let $t_k$ and $t_{k+1} \approx t_k + \period$, \review{where $\period \approx 6.55$ \SI{}{days} is the approximate orbital period of the NRHO,} denote two consecutive control epochs, such that $\theta(t_k) = \theta(t_{k+1}) = \theta_{\rm man}$. 
In each revolution, we consider one post-maneuver tracking window starting $12$ hours after $t_k$, and three pre-maneuver tracking windows, starting $72$, $48$, and $7$ hours before $t_{k+1}$. 
During each tracking window, we provide $N_{\rm meas} = 10$ equally spaced measurements. 
At $t_{k+1}$, Algorithm~\ref{alg:mpc_tracking} uses the EKF's predicted state estimate following the latest measurement update provided $t_{k+1} - 7 + \Delta t_{\rm track} = 6$ hours earlier.

\subsection{Control Trigger Condition}
\ACCrev{To improve the delta-V performance of the SKMPC under navigation and execution errors, we consider a trigger condition to determine whether a maneuver is required.}
This condition checks if the state propagated forward without control until $t_N$ lies within an ellipsoid about the baseline with radii $\epsilon_{r,\mathrm{trig}}$ in position and $\epsilon_{v,\mathrm{trig}}$ in velocity components,
\begin{align}
    \nonumber 
    \| \bar{\rbold}_{N-1|k} - {\rbold}_{f,\mathrm{ref}} \|_2 \leq \epsilon_{r,\mathrm{trig}}, 
    \, 
    \| \bar{\vbold}_{N-1|k} - {\vbold}_{f,\mathrm{ref}} \|_2 \leq \epsilon_{v,\mathrm{trig}}
    ,
\end{align}
where $\bar{\xbold}(t_{N-1|k}) = [\bar{\rbold}_{N-1|k}^T,\bar{\vbold}_{N-1|k}^T]^T$.
Tolerances $\epsilon_{r,\mathrm{trig}}$ and $\epsilon_{v,\mathrm{trig}}$ do not need to be the same as $\epsilon_r$ and $\epsilon_v$ in~\eqref{eq:termnal_set_ellipsoid}. 
Choosing $\epsilon_{r/v} < \epsilon_{r/v,\mathrm{trig}}$ makes the closed loop more robust against uncertainties; \review{recursive feasibility is recovered by modifying the proof that is based on $\epsilon_{r,\mathrm{trig}}$ and $\epsilon_{v,\mathrm{trig}}$ instead of $\epsilon_{r}$ and $\epsilon_{v}$, still assuming a sufficiently large $u_{\max}$.}

\section{Numerical Results}
\label{sec:numerical_results}
We conduct a Monte-Carlo simulation, where each sample consists of navigating and performing SK over $300$ revolutions along the baseline NRHO of the Gateway generated by NASA~\cite{Lee2019}, corresponding to over 5.3 years.
\review{We compare the performance of SKMPC against XAC.}
With the SKMPC, we use $N_{\rm rev}=8$ and $u_{\max} = 1\,\mathrm{m/s}$, with triggering thresholds $\epsilon_{r,\rm trig} = 100\,\mathrm{km}$ and $\epsilon_{v,\rm trig} = 20\,\mathrm{m/s}$, and terminal constraint radii $\epsilon_r = 25\,\mathrm{km}$ and $\epsilon_v = 5\,\mathrm{m/s}$.
\review{With XAC, we use $N_{\rm rev}=7$, i.e., targeting the $7^{\rm th}$ perilune downstream, with $\epsilon_{v_x} = 1\,\mathrm{m/s}$.
From preliminary experiments, we note that we are able to use a larger $N_{\rm rev}$ with the SKMPC due to (1) the multiple-shooting formulation and (2) reduced sensitivity targeting conditions around apolune, compared to XAC, which is a single-shooting formulation that targets conditions around perilune.}
All thresholds are defined in $\mathcal{F}_{\rm EM}$. 
We conduct three cases using 1, 2, and 3 desaturation events per revolution, at $\theta_{\rm desat}$ provided in Table~\ref{tab:error_parameters}.

\review{The dynamics is integrated using the explicit Runge-Kutta Prince-Dormand (8,9) method from the GNU Scientific Library~\cite{gough2009gnu}, rescaling~\eqref{eq:nonlinear_dynamics} with canonical distance unit $\mathrm{DU} = 10^5$~\SI{}{km} and velocity unit $\mathrm{VU} \triangleq \sqrt{\mu/\mathrm{DU}}$ to improve numerical conditioning, with relative and absolute tolerances set to $10^{-14}$.
The SKMPC takes an average of \SI{2.57}{sec} to solve on a single Intel i7-12700 CPU; the majority of the computational effort comes from propagating the STMs.}

\subsection{Navigation Performance}
\begin{figure}[t]
    \centering
    \includegraphics[width=0.9990\linewidth]{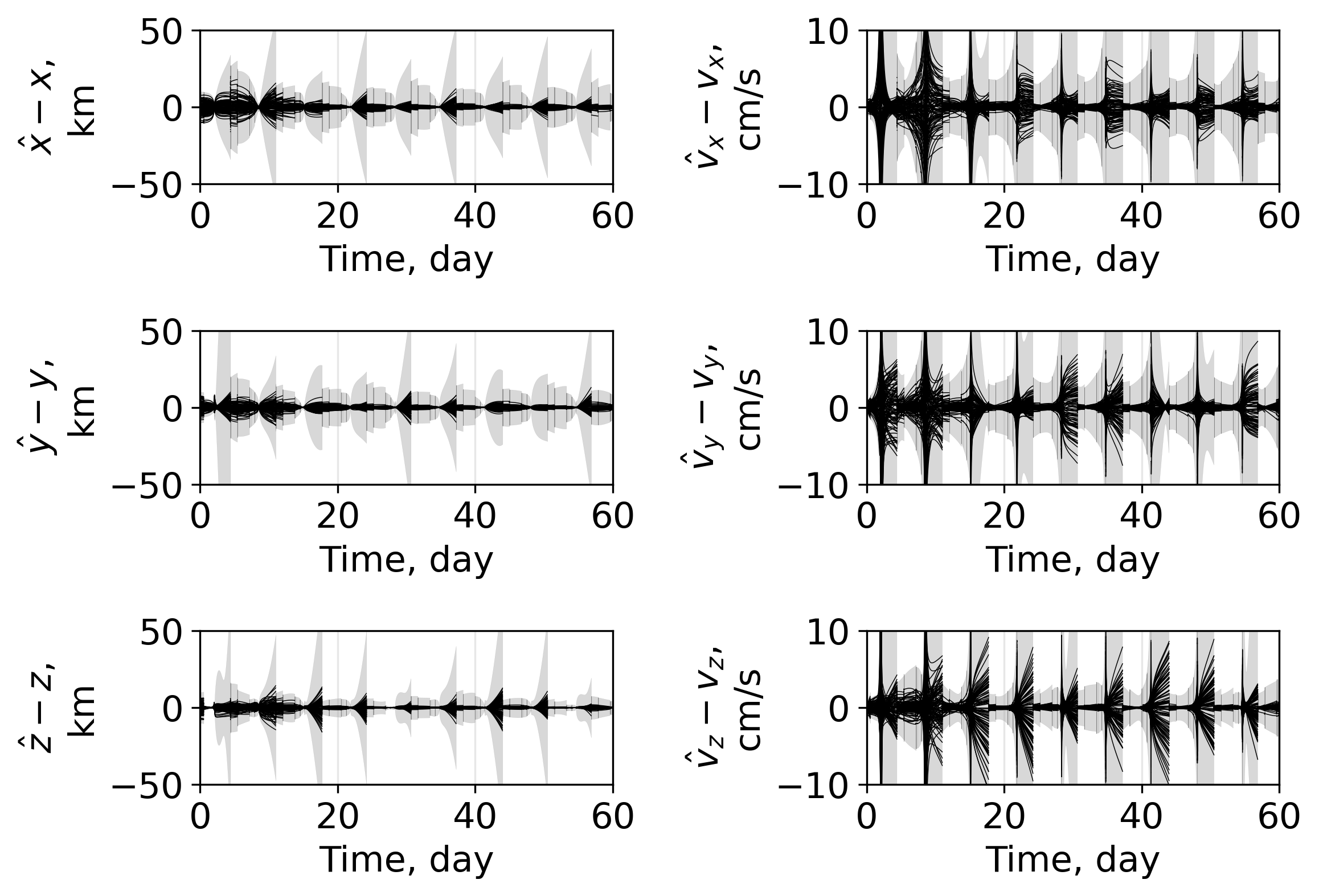}
    \caption{Estimation error in $\mathcal{F}_{\rm Inr}$ with 3 desaturation events}
    \label{fig:filter_recurse}
\end{figure}

We first look at the navigation estimates provided to the controller. 
Figure~\ref{fig:filter_recurse} shows the estimation error of the EKF for the case involving 3 desaturation events and controlled by SKMPC; only the first 60 days are shown for the sake of clarity, as the filter performance is qualitatively similar across the remaining 240 days. 
The filter performance for 1 and 2 desaturation events, \review{controlled by either the SKMPC or XAC,} is qualitatively similar. 
For assessing SK activities, we focus on navigation performance at the maneuver time. 
Table~\ref{tab:premaneuver_state_estim_error} shows the numerical $3$-$\sigma$ pre-maneuver state estimation error with 1, 2, and 3 desaturation events \review{using the SKMPC; errors using XAC are at similar orders of magnitude}.
As the number of desaturation events increases, the navigation error at the control epoch gets worse.
The spikes in velocity estimate errors and corresponding covariance correspond to perilune passes, where the spacecraft undergoes a rapid change in direction of motion as it flies by the Moon.

\subsection{Cost Performance}
\begin{figure}[ht!]
     \centering
     \begin{subfigure}[b]{0.44\textwidth}
         \centering
         \includegraphics[width=\textwidth]{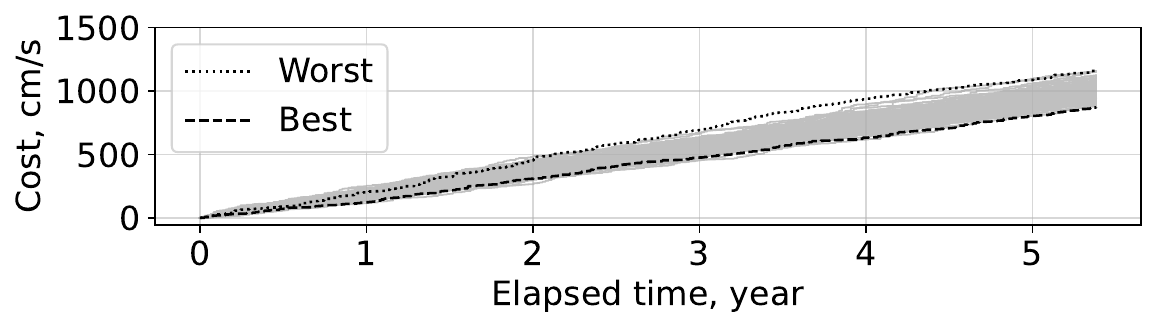}
         \caption{SKMPC, $N_{\rm rev}=8$}
         \label{fig:MC_cumulative_cost_history_SKMPC}
     \end{subfigure}
     \\[0.5em]
     \begin{subfigure}[b]{0.44\textwidth}
         \centering
         \includegraphics[width=\textwidth]{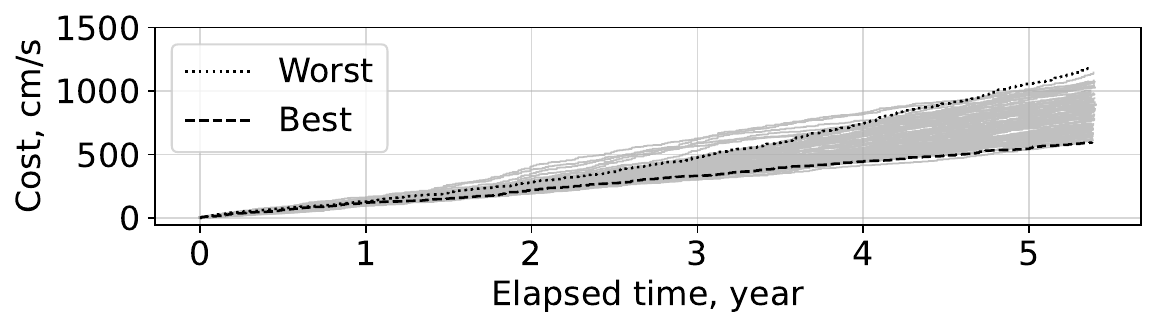}
         \caption{$x$-axis crossing control (XAC), $N_{\rm rev}=7$}
         \label{fig:MC_cumulative_cost_history_XAC}
     \end{subfigure}
    \caption{Cumulative cost history with 3 desaturation events}
    \label{fig:MC_cumulative_cost_history}
\end{figure}

Tables~\ref{tab:SKMPC_performances} and~\ref{tab:XAC_performances} provide cost statistics using the SKMPC \review{and XAC.
With both controllers,} the SK cost increases as the navigation performance worsens; across varying numbers of desaturation events, the SKMPC provides both a lower yearly cost mean and standard deviation compared to XAC.
The cumulative cost histories with 3 desaturation events per revolution are shown in Figures~\ref{fig:MC_cumulative_cost_history_SKMPC} and~\ref{fig:MC_cumulative_cost_history_XAC}.
With the SKMPC, the cumulative cost history follows a predominantly linear trend, indicating the controller is applying the appropriate level of control effort to keep the spacecraft motion near the baseline despite the uncertainties.
\review{In contrast, with XAC, the growth of the cumulative cost varies significantly based on the Monte-Carlo sample, which is a sign of the controller's susceptibility to uncertainties that lead to phase deviation.}

\begin{table}[ht!]
    \centering
    \caption{Pre-maneuver state estimation error in $\mathcal{F}_{\rm EM}$}
    \label{tab:premaneuver_state_estim_error}
    \begin{tabular}{@{}lrrr@{}}
        \toprule
        Number of desaturation events & $1$ & $2$ & $3$ \\
        \midrule
        $(\hat{x} - x)$ 3-$\sigma$, \SI{}{km}       & 0.924 & 1.041 & 1.128 \\
        $(\hat{y} - y)$ 3-$\sigma$, \SI{}{km}       & 1.068 & 1.311 & 1.492 \\
        $(\hat{z} - z)$ 3-$\sigma$, \SI{}{km}       & 0.635 & 0.677 & 0.711 \\
        $(\hat{v}_x - v_x)$ 3-$\sigma$, \SI{}{cm/s} & 0.213 & 0.222 & 0.228 \\
        $(\hat{v}_y - v_y)$ 3-$\sigma$, \SI{}{cm/s} & 0.700 & 0.927 & 1.086 \\
        $(\hat{v}_z - v_z)$ 3-$\sigma$, \SI{}{cm/s} & 0.101 & 0.119 & 0.133 \\
        \bottomrule
    \end{tabular}
\end{table}

\begin{table}[ht!]
    \centering
    \caption{Cost statistics using SKMPC}
    \begin{tabular}{lrrr}
    \toprule
    Number of desaturation events & $1$ & $2$ & $3$ \\
    \midrule
    Yearly mean, \SI{}{cm/s}                    & 109.96 & 153.48 & 186.83     \\
    Yearly standard deviation, \SI{}{cm/s}      & 8.24   &  9.99  & 12.35      \\
    Yearly $95^{\rm th}$ percentile, \SI{}{cm/s}& 123.21 & 169.34 & 208.63     \\
    \bottomrule
    \end{tabular}
     \label{tab:SKMPC_performances}
\end{table}

\begin{table}[ht!]
    \centering
    \caption{Cost statistics using $x$-axis crossing control}
    \begin{tabular}{lrrr}
    \toprule
    Number of desaturation events & $1$ & $2$ & $3$ \\
    \midrule
    Yearly mean, \SI{}{cm/s}                    & 111.94 & 135.76 & 155.29 \\
    Yearly standard deviation, \SI{}{cm/s}      & 33.26  & 30.77  & 26.94  \\
    Yearly $95^{\rm th}$ percentile, \SI{}{cm/s}& 178.78 & 192.55 & 198.66 \\
    \bottomrule
    \end{tabular}
     \label{tab:XAC_performances}
\end{table}

\subsection{Tracking Performance}
\begin{figure}[ht!]
     \centering
     \begin{subfigure}[b]{0.490\textwidth}
         \centering
         \includegraphics[width=\textwidth]{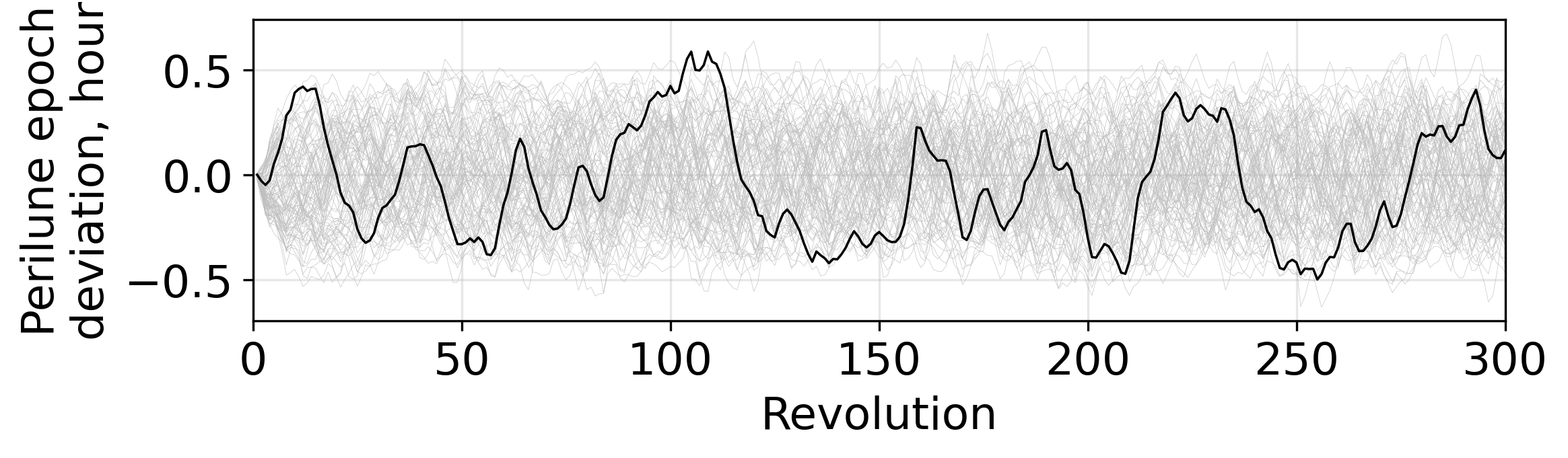}
         \caption{SKMPC, $N_{\rm rev}=8$}
         \label{fig:ALSCP_perilune_deviation_epoch_target8_horizon300_cutoff6_desat3}
     \end{subfigure}
     \\
     \begin{subfigure}[b]{0.490\textwidth}
         \centering
         \includegraphics[width=\textwidth]{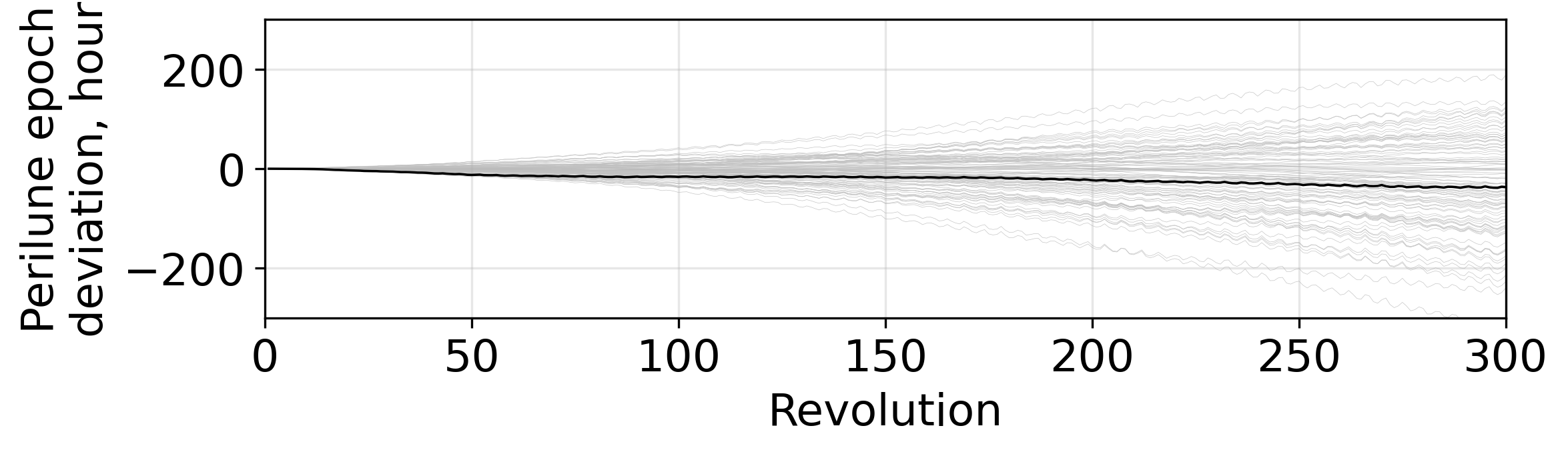}
         \caption{$x$-axis crossing control (XAC), $N_{\rm rev}=7$}
         \label{fig:XAC_perilune_deviation_epoch_target7_horizon300_cutoff6_desat3}
     \end{subfigure}
    \caption{Epoch deviation at perilune passes}
    \label{fig:perilune_epoch_deviation}
\end{figure}

\begin{figure}[ht!]
     \centering
     \begin{subfigure}[b]{0.490\textwidth}
         \centering
         \includegraphics[width=\textwidth]{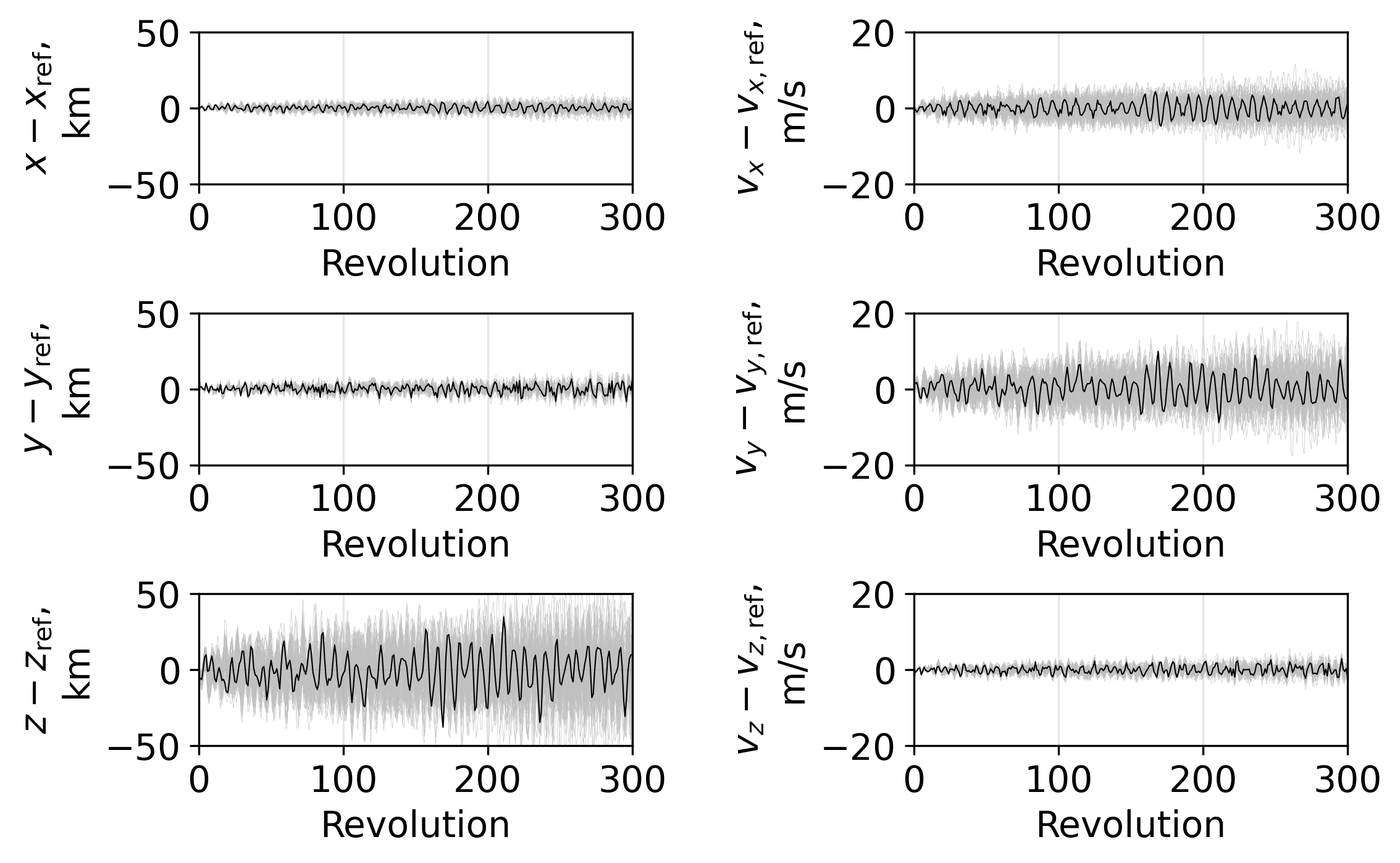}
         \caption{SKMPC, $N_{\rm rev}=8$}
         \label{fig:ALSCP_perilune_deviation_state_target8_horizon300_cutoff6_desat3}
     \end{subfigure}
     \\[0.5em]
     \begin{subfigure}[b]{0.490\textwidth}
         \centering
         \includegraphics[width=\textwidth]{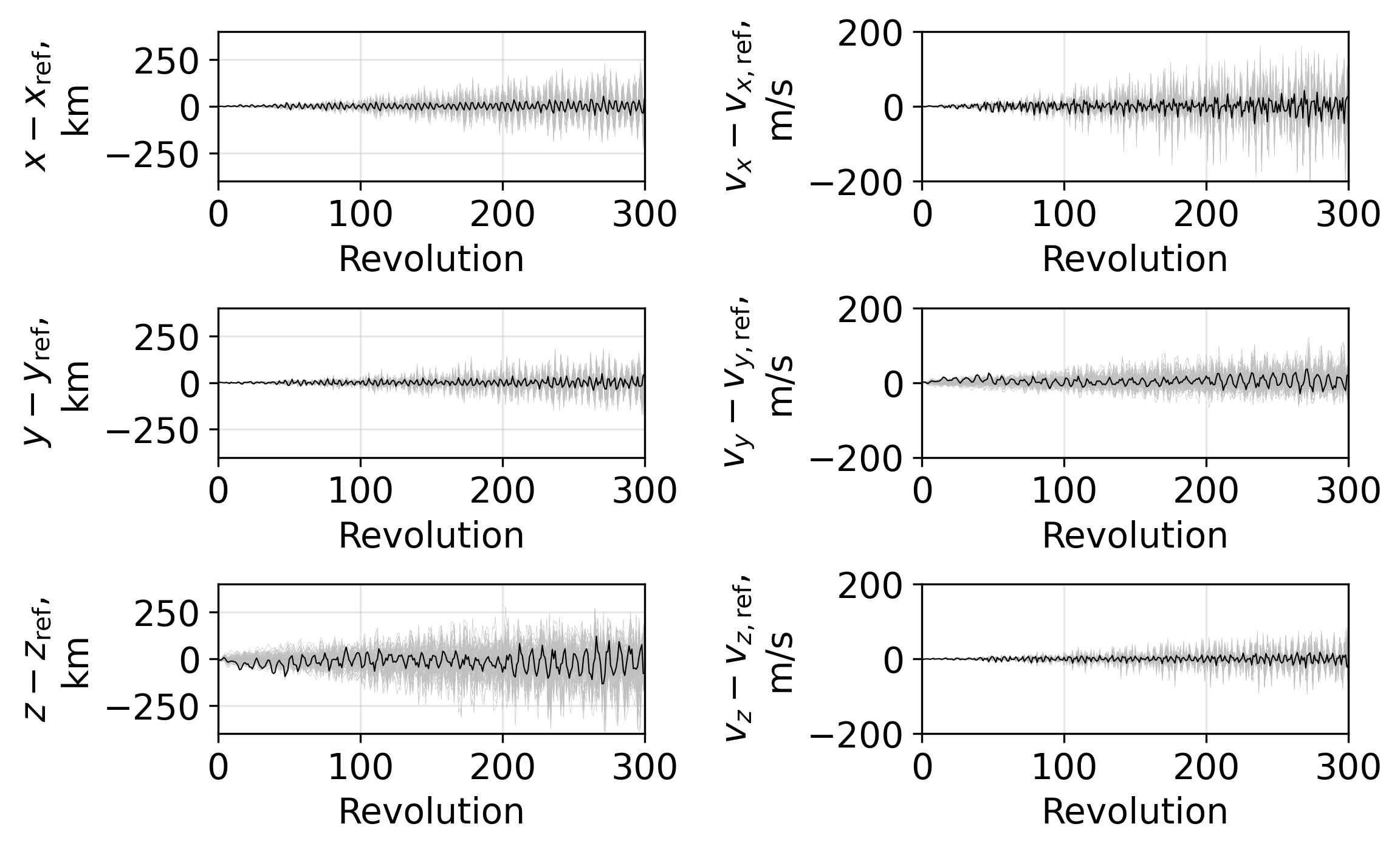}
         \caption{$x$-axis crossing control (XAC), $N_{\rm rev}=7$}
         \label{fig:XAC_perilune_deviation_state_target7_horizon300_cutoff6_desat3}
     \end{subfigure}
    \caption{State deviation in $\mathcal{F}_{\rm EM}$ at perilune passes}
    \label{fig:perilune_state_deviation}
\end{figure}

We now analyze the tracking capability of the SKMPC \review{and XAC}.
To isolate the effect of phase deviation, we compare the epochs and states at perilune passes of the controlled trajectory to the corresponding perilune passes of the baseline. 
Figures~\ref{fig:perilune_epoch_deviation} and~\ref{fig:perilune_state_deviation} show the deviation of the epoch and state in $\mathcal{F}_{\rm EM}$ at each perilune passage.
Note that in $\mathcal{F}_{\rm EM}$, perilunes occur approximately along the $+z$ axis, with the spacecraft's motion approximately perpendicular to the position vector; thus, the error is expected to be larger in $z$ compared to $x$ and $y$ in position components, and $v_x$, and $v_y$ compared to $v_z$ in velocity components.

\review{XAC experiences secular growth in phase deviations; the epoch deviation at perilune, while varying largely on the Monte-Carlo realization, reaches the order of several days, with passes deviating by up to \SI{500}{km} in position and \SI{200}{m/s} in velocity.
By augmenting XAC with residual on perilune epoch deviation, Davis et al.~\cite{Davis2022} reports deviations of up to \SI{80}{km} in position and $48$ minutes in epoch.}

\review{In contrast, through its full-state tracking property, the SKMPC keeps the perilune epoch deviation to about $30$ minutes, with position deviations of about \SI{50}{km} and velocity deviations of about \SI{10}{m/s}.}
In general, tighter tracking of the baseline is desirable since more stringent requirements can be met with regard to the spacecraft design or payload operations that require remaining closer to the intended path. 
For instance, the NRHO baseline for the Gateway is designed to be free of any Earth-shadowing eclipses~\cite{ZimovanSpreen2022}, and tight tracking despite the dynamical uncertainties and errors can ensure that no such eclipse occurs during the flight. 

\section{Conclusion}
\label{sec:conclusion}
We proposed an economic MPC for the SK problem on the NRHO. 
This SKMPC achieves full-state tracking by taking into account multiple maneuvers within its control horizon. 
The controls are placed one revolution apart, making our approach \review{operationally compliant} to the single maneuver-per-revolution requirement typical in space missions on the NRHO. 
Through full-state tracking, the SKMPC overcomes the issue of \review{phase deviation} encountered by other state-of-the-art SK schemes with a single maneuver per revolution.
We tested the SKMPC with output-feedback using an EKF with range and range-rate measurements, and subject to realistic error models.
Our approach achieves cumulative maneuver costs comparable to SK approaches proposed in the astrodynamics literature, while resulting in tighter tracking of the baseline orbit in both space and phase without additional ad-hoc heuristics, as required in XAC. 
\balance

\bibliographystyle{IEEEtran}
\bibliography{IEEEabrv,references.bib}
\end{document}